\documentclass[twocolumn]{article}
\usepackage{graphicx}
\usepackage{amsmath}
\usepackage[style=numeric,sorting=none]{biblatex}
\usepackage{comment}

\title{The identification of garbage dumps in the rural areas of Cyprus through the application of deep learning  to satellite imagery}
\author{
    Andrew K. Wilkinson \\
    University of York \\
    \texttt{akw314159265@gmail.com}
}
\date{\today}

\bibliography{bibliography}

\begin{document}

\maketitle

\begin{abstract}

Garbage disposal is a challenging problem throughout the developed world.  In Cyprus, as elsewhere, illegal ``fly-tipping" is a significant issue, especially in rural areas where few legal garbage disposal options exist.  However, there is a lack of studies that attempt to measure the scale of this problem, and few resources available to address it.  A method of automating the process of identifying garbage dumps would help counter this and provide information to the relevant authorities.

The aim of this study was to investigate the degree to which artificial intelligence techniques, together with satellite imagery, can be used to identify illegal garbage dumps in the rural areas of Cyprus.  This involved collecting a novel dataset of images that could be categorised as either containing, or not containing, garbage. The collection of such datasets in sufficient raw quantities is time consuming and costly.  Therefore a relatively modest baseline set of images was collected, then data augmentation techniques used to increase the size of this dataset to a point where useful machine learning could occur.

From this set of images an artificial neural network was trained to recognise the presence or absence of garbage in new images.  A type of neural network especially suited to this task known as ``convolutional neural networks" was used.  The efficacy of the resulting model was evaluated using an independently collected dataset of test images. 

The result was a deep learning model that could correctly identify images containing garbage in approximately 90\% of cases.  It is envisaged that this model could form the basis of a future system that could systematically analyse the entire landscape of Cyprus to build a comprehensive ``garbage" map of the island.

Keywords: garbage disposal, fly-tipping, artificial intelligence, satellite imagery, image augmentation, convolutional neural networks, deep learning, Cyprus.

\end{abstract}

\section{Introduction} \label{intro}

\subsection{Background and Motivations}

Illegal waste dumping is a social and environmental issue throughout the developed and developing world \cite{vrijheid_health_2000}. The consequences include environmental pollution, health hazards, and negative impacts on local ecosystems and aesthetics. The island of Cyprus has particular issues with waste disposal in rural areas\cite{gabrielides_man-made_1991}, where little formal recycling exists.  As a result, citizens tend to make impromptu garbage dumps. Assessing the scale of this issue is difficult \cite{webb_fly-tipping_2006}.  However, if there was a way of identifying how many such garbage dumps have accumulated, together with their location, then this could be used to inform and exert pressure on local, regional and national government bodies to address the issue proactively \cite{kinnaman_garbage_1997}.

Performing this sort of task by hand would be laborious and error prone.  If, for example, the land was divided up into $100m^{2}$ patches, then even for a relatively modest sized country such as Cyprus covering around $9250km^{2}$ this would require at least 925,000 such image patches to be analysed.  If a human could retrieve an image, visually analyse it, and store the classification result as an atomic operation in around one minute, then to classify all image patches would require more than 15,400 person hours or 640 days.

Some previous work has employed AI techniques to automatically detect the presence of garbage dumps, such as \cite{rajkumar_detecting_2022} and \cite{devesa_mapping_2021}.  These projects have focused on large land-fills occupying an area of several square kilometers.  Few such projects exist focusing on small scale sites covering just tens or hundreds of square metres.  However, it is such small areas that Cyprus garbage dumps tend to occupy.  With the existence of public access high resolution remote sensing imagery, such as satellite \cite{wulder_current_2019}  \cite{zhao_progress_2021}, it is possible to focus on techniques to recognise these smaller scale features.  This study utilised a type of Artificial Neural Network (ANN) shown to be effective in the classification of satellite imagery \cite{patowary_lightweight_2020} and applied it to these small scale features.

\subsection{Aims and Objectives} \label{rqs}

The aim of this study was to develop and evaluate an AI based approach for identifying garbage heaps in Cyprus.  This involved the collection and labeling of satellite imagery.  However, collection of such image sets is a laborious task and it is unfeasible to collect such ``raw" data in sufficient quantities to perform useful machine learning \cite{ghaffar_data_2019}.  To counter this, data augmentation techniques such as sharpening, cropping, rotating, and flipping were used to significantly increase the dataset sizes, whilst retaining the important features.

The questions the study sought answers to were:
\begin{itemize}
    \item
What are the most effective machine learning algorithms and techniques for developing a model that can accurately identify and classify small scale garbage dumps in the rural areas of Cyprus from satellite imagery?
\item 
To what extent can data augmentation techniques improve a machine learning model for such small scale satellite image classification?
\end{itemize}

\subsection{Approach}
To provide answers to the questions above the following steps were completed.
Data was collected in the form of satellite images of known garbage locations in Cyprus, forming a baseline dataset. This dataset was expanded through various data augmentation techniques \cite{shorten_survey_2019}: sharpening, rotating, cropping, and flipping.  A suitable machine learning model was selected for the study by training and evaluating various candidates against the baseline. The chosen model was then trained on the augmented datasets.  Finally, the accuracy of the model was evaluated by applying various validation methods and statistics in order to arrive at conclusions with regards the efficacy of the model.

The study focused on a type of deep learning model referred to as convolutional neural networks (CNNs).  These work by applying various filters to the images which learn to pick out low level then progressively higher level features from the images \cite{simonyan_very_2014}, culminating in, for this study, the identification of garbage.  Dimensionality of the input data is reduced to a manageable level by use of such techniques as pooling \cite{shamsolmoali_high-dimensional_2019}.  CNNs have proven effective at image recognition, and have been applied successfully to the area of remote sensing \cite{castelluccio_land_2015}.

\subsection{Results and Contributions}

The results and outputs of the study are manyfold:
\begin{itemize}
    \item A novel baseline set of satellite image patches labelled as containing/not containing garbage, which might find use outside of this study.
    \item Several augmented datasets of similarly labelled images which may also find wider use within the machine learning community.
   \item A set of CNN models which provide an accurate basis for garbage heap identification with the potential to inform waste management policies in Cyprus.
\end{itemize}

\subsection{Outline}
Section \ref{lit_rev} develops a theoretical framework to the study, mostly through a review of pertinent literature. Section \ref{rm} establishes the research methodology used to answer the research questions.  This is followed by a discussion of the results in section \ref{res} and finally a conclusion and recommendations for future work in section \ref{conc}.

\section{Theoretical Background}  \label{lit_rev}

\subsection{Motivations} \label{lit_rev_mot}
Webb et al., 2006 \cite{webb_fly-tipping_2006} provide a comprehensive study into the incidence of fly-tipping.  They describe in detail the who, how, where, and why of such practices and their environmental and social impact. One issue highlighted is the difficulty authorities have in mapping the location of fly-tipping sites - a main driver behind this research and one of the main objectives it seeks to address by producing models which can recognize garbage dumps from satellite images.

\subsection{Deep Learning}

The original deep learning research into image recognition was undertaken thirty years ago when LeCun used the Multilayer Perceptron with back-propagation to recognize handwritten digits \cite{lecun_backpropagation_1989}.  Back-propagation is an iterative algorithm used to adjust the weights of the connections between nodes to minimize the difference between the predicted output and the actual output.  It forms the basis by which Artificial Neural Networks (ANNs) are trained on input data.
Since then deep learning has become the dominant method for all image recognition tasks, including those involving remote sensing sources such as satellite imagery. One of the main reasons for this is that an ANN can model complex non-linear relationships, as described by LeCun in the later 2015 paper \cite{lecun_deep_2015}.  

Standard ANNs can be effective for classifying small image patches, but for complex large satellite images this is an impractical method.  Each pixel of an image is a separate input to the network.  Once multiple layers are added to the network then the number of parameters in the training process grows exponentially such that the training time and memory requirements become prohibitively high \cite{charu_c_aggarwal_basic_2018}.  To address this a type of ANN called the Convolutional Neural Network (CNN) has become very popular. 

\subsection{Convolutional Neural Networks} \label{lit_rev_cnn}

The first CNN was proposed by Fukushima in the foundational 1980 paper \cite{fukushima_neocognitron_1980} in which the neocognitron is described as an ``improved neural network".  This is essentially a hierarchy of networks composed of layers of nodes with ``variable connections between adjoining layers". LeCun took this idea further in 1998 \cite{lecun_gradient-based_1998} in which back-propagation and gradient descent were added and he showed how these strategies could solve pattern recognition involving high dimensional images better than existing hand-designed algorithms.
This lay the groundwork for future CNN algorithms and variations such as LeNet, AlexNet, Google LeNet, and ResNet. \cite{alom_history_2018}.  Much of the earlier work focused on recognition of small scale images, particularly that of handwritten characters \cite{baldominos_survey_2019}.  Subsequently, this was adapted for larger scale satellite imagery.

\subsection{Machine Learning and Satellite Imagery}  \label{lit_rev_ml}
Satellite imagery poses unique challenges for machine learning. Forkuor et al. (2018) extensively discuss these challenges in the context of land-use and land-cover mapping in Burkina Faso. Satellite images are larger and multi-spectral, with Sentinel-2 satellites offering 13 spectral bands, including near-infrared channels. This additional spectral information enables analysis of thermal characteristics, which can be relevant for certain types of garbage dumps containing biological matter. However, utilizing more channels in training a model increases the number of parameters and learning time. Consequently, this study opted to use three-channel RGB images.

Vaishnnave et al. (2019) provide a survey of different methods for classifying satellite datasets, including popular CNN algorithms like AlexNet, ResNet-50, GoogleNet, and CaffeNet. CNN methods exhibited the highest accuracy, ranging from 93\% to 99\%. It is important to note that these accuracy figures are drawn from various studies with different datasets and objectives, making direct comparisons challenging. Nonetheless, they lend support to the approach adopted in the current study.

A particularly relevant study by Rajkumar et al. (2022) applies various ML algorithms to landfills using satellite missions such as WorldView and GeoEye. The dataset comprised 245 images, each with dimensions of 512x512 pixels. Although the covered area is not specified, the landfills were significantly larger than those addressed in the current study. The results demonstrated accuracy ranging from 76\% to 83\%. 

\subsection{Residual Networks} \label{lr:resnet}
Residual Networks (ResNets) are a successful subset of CNNs, particularly in remote sensing. He et al. introduced ResNets in 2016 to address the vanishing gradient problem, which hinders learning in deep networks. The influential paper proposes an architecture with shortcut connections between layers, allowing nodes to be treated as blocks during the learning process. This approach facilitates the smooth flow of gradients through deep networks. The paper demonstrates the capability of training networks with over 1000 layers using this method.

\subsection{Pretrained Zoo Models} \label{lr:pret}
Pretrained models are specific implementations of ML algorithms that have been trained on large amounts of data. They can be used to ``shortcut" the training time of similar problems and result in better models. The paper by Yosinski et al. 2014 \cite{yosinski_how_2014} details early work that showed how even when the training data is significantly different from that of the target problem, that models designed around them perform better than not pretrained. It is not fully established why this is, but it might have something to do with the general structures that make up typical objects in 2D images being learned, such as edges, irrespective of what the target image is.

\subsection{Data Collection} \label{lit_rev_dc}
There are several potential sources of high resolution satellite data. 
The US geological survey provides Earth Explorer - a user interface into the Landsat satellite system \cite{united_states_geological_survey_usgs_nodate}.  Sentinel data is available from the ESA open access hub \cite{european_space_agency_copernicus_nodate}. Forkuor (2018)  \cite{forkuor_landsat-8_2018} describes a study of these two sources for land use and land cover, and applied three ML algorithms to compare their relative performance.  Of the two, the Sentinel data
proved slightly more accurate than the Landsat, although the algorithms were not ANN based, which tempers the applicability of the results.

An additional potential source of images is the Google Earth engine \cite{google_corp_google_nodate}. This provides a browser front-end and API into Landsat and Sentinel imagery, which are freely available for third-party use.  Zhao et al. 2021 \cite{zhao_progress_2021} summarises the Google Earth engine and categorises the types of use to which the engine has been applied, including land classification.

\subsection{Data Augmentation} \label{lit_rev_da}
Obtaining labelled satellite imagery for most ML is difficult and time consuming \cite{ghaffar_data_2019}.  It can be especially challenging to source images in sufficient numbers to obtain meaningful results.  One way to address this is to use image augmentation techniques to enlarge the base set of images. Shorten et al. 2019 \cite{shorten_survey_2019} offers a comprehensive overview of the various approaches, including geometric and photometric.  Geometric techniques focus on changing an image by rotating, cropping, and scaling.  Photometric focuses on changing an image by altering the contrast, sharpening, and brightness. 

These two approaches are studied in more detail by Taylor et al. (2018) \cite{taylor_improving_2018} in which the results of performing ML experiments on both are described.  The DeepLearning4J framework \cite{lang_wekadeeplearning4j_2019} and CNNs were used together with a variety of augmentation techniques and a comparison made. The study does not record the results of applying all techniques together at once as ``pipelines", but does suggest that geometric cropping provides the biggest improvement from 48.13\% to 61.95\%.

\section{Research Methodology} \label{rm}

\subsection{Research Design} \label{rm_rd}

In order to achieve the aims and objectives introduced in section \ref{rqs} the following approach was adopted:

\begin{itemize}
\item To collect a set of satellite images from a publicly available source.
\item To appropriately label each image as either ``garbage" or ``not garbage".
\item To apply several promising supervised deep learning approaches to this baseline set of images and to choose the best performing for the remainder of the study.
\item To apply various data augmentation techniques to the baseline dataset in order to derive enlarged datasets.
\item To evaluate each enlarged dataset  against the best performing machine learning model and make appropriate performance comparisons.
\end{itemize}

The sections that follow describe the above process in detail together with limitations of the study.

\subsection{Data Collection} \label{rm_dc}

The data collected for this study was RGB colour satellite images centred on latitude/longitude  coordinates known to contain garbage dumps.  These points were collected over the last year and are taken from a variety of rural locations.  They provide a representative set of images of the sort of garbage dumps found throughout Cyprus, taken from a variety of land types, such as forest, scrubland, mountain, and arable. 

For every image collected of a garbage dump site, a corresponding image of a nearby location that did not contain garbage was collected; this ensured a balanced dataset of both classes of interest (garbage, not-garbage).  A balanced dataset is important to prevent biases in the learned model, as discussed in \cite{noauthor_handling_nodate}.  A baseline set of 100 images was collected - 50 labelled as ``garbage" and 50 labelled as ``not garbage".

Various possible sources of satellite imagery exist, as detailed in \ref{lit_rev_dc}.  It was decided for this study that the best source would be Google Earth \cite{google_corp_google_nodate}.  This decision was made due to its ease of use and availability, and because it covers the areas of interest with little cloud cover which would otherwise obscure the features of interest.  Google earth has also been used successfully on various other similar land classification projects, such as \cite{devesa_mapping_2021} and \cite{skogsmo_scalable_2020}. 

The images collected were 200x200x3 files and upscaled as necessary for the model.  These were a convenient size to work with and allowed the features of interest to be placed within a single image per occurrence.  

\begin{figure}[!ht]
\includegraphics[scale=0.58]{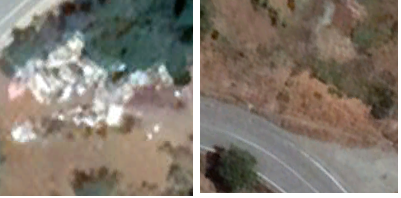}
\vspace{-0.5em}
\caption{Garbage/Not garbage}
\label{fig:g_big}
\end{figure}

Figure \ref{fig:g_big} illustrates a collected garbage/not garbage image pair.

Each 200x200 image maps onto a land area of around 20x20m area, with one pixel representing around 10x10cm.

\subsection{Data Augmentation} \label{rm_da}

Labelled satellite imagery for ML projects of this sort is laborious to collect in large enough quantities to produce accurate and robust models on their own \cite{basu_deepsat_2015}.  Studies have frequently relied on image augmentation to enlarge the dataset whilst preserving the labelling characteristics of the images. Section \ref{lit_rev_da} reviewed methods of increasing an image set set via image augmentation.

The geometric augmentations most likely to offer improvement according to \cite{taylor_improving_2018} were applied as follows (inclusive of the original image):

\begin{itemize}
    \item cropping: each suitable garbage labelled image and their non-garbage counterpart was subject to the cropping method described in the previous section. 
    \item rotating: each image was rotated by 90, 180, and 270 degrees. 
    \item flipping: each image was flipped horizontally and vertically. 
\end{itemize}

The most promising photometric technique seemed likely to be sharpening, which can improve and enhance edge-detection capabilities, as outlined in \cite{ghaffar_data_2019}.  The images were all subjected to a sharpening ``kernel" filter.

By combining these augmentation approaches in various ways several datasets were created, ranging from 100 images for the baseline set up to 2,400 images for a pipelined dataset combining all of the techniques above.

\subsection{Machine Learning}
Once suitably augmented sets of data were obtained, the ML process was carried out  as follows:

\begin{itemize}
    \item choose model instance to go forward with
    \item tune hyperparameters
    \item run model against each augmented dataset
    \item run model against pipelined datasets
    \item evaluate the results
\end{itemize}

This iterative process of tuning hyperparameters, training on augmented datasets, combining them in pipelined datasets, and evaluating the results helped in improving the model's performance and ability to handle novel scenarios.

\subsubsection{Choosing the model} \label{meth:choose}
Only convolutional neural network (CNN) models were considered for the actual machine learning, as these have been found to be most effective for the sort of problem addressed by this study \cite{vaishnnave_study_2019}, as explored in section \ref{lit_rev_cnn}.

There are many possible CNN models to choose from.  It was decided to evaluate a range of those that had been successfully applied to image classification tasks, including several winners of the ImageNet Large Scale Visual Recognition Challenge (ILSVRC) \cite{russakovsky_imagenet_2015}.
The process was to apply each candidate model to the baseline (unaugmented) and compare the relative performance of the model, using the metrics described in section \ref{res:stats}.
\cite{rajkumar_detecting_2022} and \cite{vaishnnave_study_2019} suggest that Residual Network model might be most effective when applied to this dataset, which proved to be the case. The results section \ref{res:mod_sel} summarises the outcome of this model selection process, including a list of the candidate models evaluated.  

All the models evaluated were pretrained.  These have already been trained against an image dataset, in this case the publicly available ImageNet dataset \cite{stanford_vision_labs_imagenet_2020}.  Whilst not consisting exclusively of satellite imagery (in fact only a small proportion is) it has been found that even models pretrained on non-satellite images have an advantage over those not pretrained at all \cite{yosinski_how_2014}.

\subsubsection{Modelling Individual Datasets} \label{meth:single}
The following datasets were modelled to establish the relative effectiveness of each data augmentation technique via the selected model, as determined in the results, section \ref{res:mod_sel}:

\begin{itemize}
    \item baseline dataset of 100 samples
    \item geocropped dataset of 200 samples
    \item sharpened dataset of 200 samples
    \item flipped dataset of 300 samples
    \item rotated dataset of 400 samples
\end{itemize}

For each such dataset performance metrics were collected, as described in section \ref{res:stats}.  The aim of this was to establish a set of baselines for comparison of the various data augmentation techniques and for later evaluation of improvements (or otherwise) as they were pipelined together.

\subsubsection{Modelling Pipelined Datasets} \label{meth:pipelined}

There are numerous ways of combining the individual datasets, and it is only practicable to model a subset of these.  For this study, to give a reasonable variation of datasets the following pipelines were modelled:

\begin{table}[htbp]
\caption{Pipelined Models}
  \label{tab:method:pipelining}
  \resizebox{\columnwidth}{!}{
\begin{tabular}{|p{0.12\textwidth}|p{0.68\textwidth}|p{0.12\textwidth}|}

    \hline
    \textbf{Pipeline Label} & \textbf{Description} & \textbf{\#Samples} \\
    \hline
    pipeline\_1 & Add each individual dataset into total dataset (the baseline set added in just the once)& 800\\
    \hline
    pipeline\_2 &  For each sample, rotate the sample, flip the sample, add those 6 samples to total dataset. Then sharpen the sample and rotate and flip that sharpened sample, add those 
    6 samples to the total dataset  & 1200\\
    \hline
    pipeline\_3 & Perform as per pipeline\_2 but also do the same with the cropped dataset, doubling the number of samples & 2400 \\
    \hline 
    
  \end{tabular}
}
\end{table}

Note that rotation operations multiply each sample by 3 (90°, 180°, 270°) and flipping operations multiply each by 2 (horizontal and vertically flipped) - in each case excluding the original sample (to avoid repeatedly including it in the dataset).

\subsection{Statistics Gathered} \label{res:stats}
The research undertaken represents a binary classification problem.  The model produced classifies image patches as either containing garbage (the ``positive" outcome) or not (the ``negative" outcome).  This implies that for any attempt to classify a particular image, when compared to the actual classification, there are four possibilities \cite{muller_model_2017}: 
\begin{enumerate}
    \item the classification is ``contains garbage" and the image does actually contain garbage: a ``true positive" (TP)
    \item the classification is ``contains garbage" but the image does not actually contain garbage: a ``false positive" (FP)
   \item the classification is ``does not contain garbage" and the image does not actually contain any garbage: a ``true negative" (TN)
    \item the classification is ``does not contain garbage" but the image does actually contain garbage: a ``false negative" (FN)
\end{enumerate}

Most of the statistics used to assess the efficacy of a binary classification problem make use these four quantities.  The ones used in the study, either directly or buried within another statistic, are detailed below.

\textbf{Accuracy}: the percentage of correctly classified predictions with respect to the entire dataset. This gives a good quick indication of how well the model is performing, especially for balanced datasets. See equation  \ref{eqn:acc} for how it is calculated from the figures referred to previously.

\begin{equation} \label{eqn:acc}
\mathrm{Accuracy} = \frac{\mathrm{TP} + \mathrm{TN}}{\mathrm{TP} + \mathrm{TN} + \mathrm{FP} + \mathrm{FN}}
\end{equation}

\textbf{Precision}: how well the model predicts positive results.  See equation \ref{eqn:prec}.

\begin{equation} \label{eqn:prec}
\mathrm{Precision} = \frac{\mathrm{TP}}{\mathrm{TP + FP}}
\end{equation}

\textbf{Recall}: the fraction of true positive instances that are correctly identified by the model. See equation \ref{eqn:recall}.

\begin{equation} \label{eqn:recall}
\mathrm{Recall} = \frac{\mathrm{TP}}{\mathrm{TP + FN}}
\end{equation}

These metrics give valuable insight into how well a model is performing in terms of predictive accuracy, but they do not provide a complete picture \cite{powers_evaluation_2020}.  There are various useful ways in which these statistics can be combined to produce a more complete picture.

\textbf{F-Score}: this balances precision and recall by calculating their harmonic mean.  See equation \ref{eqn:f1}.

\begin{equation} \label{eqn:f1}
\mathrm{FScore} = 2\times \frac{{\mathrm{Precision}\times\mathrm{Recall}}}{\mathrm{Precion}+\mathrm{Recall}}
\end{equation}

However, it can be seen that with the F-Score the true negative quantity is not taken into account (because neither precision nor recall do so). 

\textbf{MCC}
A statistic which takes account of all parts of the confusion matrix is the Matthew Correlation Coefficient (MCC)\cite{boaz_shmueli_matthews_2019}.  This measures the correlation between the actual and predicted classes, and yields a value between -1 and 1. An MCC of 1 indicates perfect classification, 0 random performance, and -1 a wholly negative correlation.  See equation \ref{eqn:mcc}.  This takes equal account of all four parts of the confusion matrix, so it is one of the preferred statistics used on this study.

\begin{equation} \label{eqn:mcc}
\resizebox{0.9\columnwidth}{!}{$
    \text{MCC} = \frac{(TP \times TN - FP \times FN)}{\sqrt{(TP + FP)(TP + FN)(TN + FP)(TN + FN)}}
$}
\end{equation}

\subsection{Frameworks/Tools} \label{meth:weka_etc}
The research was conducted using the Weka machine learning toolkit (Waikato Environment for Knowledge Analysis) \cite{eibe_frank_mark_a_hall_and_ian_h_witten_weka_2016}.  This is an extensible Java based toolkit that provides support for data preprocessing, implementations of many popular machine learning algorithms, visualisation tools, results statistic reporting, and much more.

For this research Weka was used in conjunction with DeepLearning4J \cite{raj_java_2019}, which provides  Java implementations of many deep learning models, including convolutional neural networks.  Various popular models are provided, such as AlexNet, LeNet, ResNet and Inception.  One especially useful feature is that many of these come as ``zoo models" \cite{shu_zoo-tuning_2021} pretrained on several publicly available image datasets, including ImageNet \cite{stanford_vision_labs_imagenet_2020}.  These pretrained models significantly shorten the training time required for new datasets \cite{yosinski_how_2014}, as detailed in section \ref{lr:pret}.

Both Weka and DeepLearning4j provide programmatic API and command line interfaces.  Weka also offers a user friendly GUI with which to supply datasets, invoke models, and visualise the results. 

In addition, LAMP was used to provide simple data collection and storage utilities \cite{yu_design_2010}.  Data augmentation was implemented using the popular ffmepg \cite{ffmpeg_team_ffmpeg_nodate} graphics and video editing software.

All development and machine learning took place on the Linux Ubuntu 20.04 operating system.
 
\subsection{Limitations} \label{meth:lims}

A CNN project requires large and diverse datasets for effective training \cite{ghaffar_data_2019}. This project has been based on a modest baseline set of 100 images. 
However, data augmentation techniques have boosted the dataset size up to 2400.  Nevertheless, additional images added to the baseset would add to the predictive performance and generalisation capabilities of the model.

The images were gathered at a resolution of 1 pixel representing 10x10cm of land coverage.  The model might not perform as well against different resolution images.
In addition, during the research the same land area was found to have  different chromatic characteristics depending on what date satellite images were taken. An area for future exploration is to examine how the developed model performs against images that differ in this way.  Possibly a model developed  against a grey scale could be more robust against such variability.  Converting the images to grey scale and rerunning the tests would be an interesting area for future study.

Many of the images were collected in mountainous wooded areas, which tend to predominate on the island. The effects of this were mitigated to an extent by using an independently collected test dataset from other areas of the island without these characteristics.  Nevertheless a larger more diverse dataset would only improve the accuracy and robustness of the model.  

It may also be the case that trees obscure some instances of garbage from satellite imagery in the visible spectrums, leading to false negatives.  It would be an interesting future study to examine if expanding the image spectrum into the infrared might alleviate this.

\section{Results and Discussion} \label{res}

This section evaluates the results of the various experiments: to first choose the best performing model type, then optimise the hyperparameters and finally run the model against the various datasets, both unaugmented and augmented.  At each stage statistics such as accuracy, F-Score and MCC were collected (see section \ref{res:stats}) and used as the basis for comparison.

\subsection{Model Selection Results} \label{res:mod_sel}

Table \ref{tab:model_eval} shows the results of evaluating the various models to determine a single candidate to move forward with, as described in section \ref{meth:choose}.  This shows the accuracy, F-Score and MCC metrics for a run of each model on the 100 image baseline dataset.

\begin{table}[htbp]
    \caption{Model Evaluation}
\label{tab:model_eval}
\resizebox{\columnwidth}{!}{
  \centering
  \begin{tabular}{|l|c|c|c|} 
    \hline
  \textbf{Model}& \textbf{Accuracy}& \textbf{F-Score}& \textbf{MCC}\\
  \hline
    LeNet & 53.00 & 0.53 & 0.06 \\
    \hline
    VGG & N/A & N/A & N/A \\
    \hline    
    AlexNet & 50.00 & 0.45 & 0.00 \\
    \hline\
    KerasInceptionV3& 47.00 & 0.47 & -0.06 \\
     \hline
    Xception & 50.00 & 0.37 & 0.00 \\
    \hline
    ResNet-50 & 67.00 & 0.66 & 0.36 \\
    \hline   
  \end{tabular}
}
\end{table}

From this it can be seen that the ResNet-50 implementation gives the best performance against all measures \footnote{The VGG implementations yielded arithmetic underflows, possibly due to the ``vanishing gradient"    problem appearing for these models \cite{hochreiter_vanishing_1998}}.
This is consistent with the results of other satellite based land classification studies, as highlighted in section \ref{lit_rev_ml}.
The ResNet-50 model was therefore selected for subsequent steps in the process. This model has 48 convolutional layers, one MaxPool layer, and one average pool layer and a total of of around 23.5 million trainable parameters \cite{he_deep_2016}.

\subsection{Singly Augmented Datasets}

Table \ref{tab:individual_eval} summarises the results of applying a single augmentation technique to the baseline dataset and adding the augmented dataset to the baseline; the ResNet-50 algorithm was applied to each of these datasets in turn. The trained model used a mini-batch size of 8 and 5x cross-fold validation as described in \cite{charu_c_aggarwal_evaluating_2018}.

\begin{table}[htbp]
\caption{Single Augmentation Evaluation}
\label{tab:individual_eval}
\resizebox{\columnwidth}{!}{
\centering
  \begin{tabular}{|l|c|c|c|c|}
    \hline
  \textbf{Dataset}& \textbf{\#samples}& \textbf{Accuracy}& \textbf{MCC}& \textbf{F-Score}\\
  \hline
    Collected baseline & 100 & 56.00 & 0.12 & 0.56\\
    \hline
    Baseline + Cropped & 200 & 83.50 & 0.68 & 0.83\\
    \hline
    Baseline + Sharpened & 200 & 74.00 & 0.05 & 0.74\\
    \hline
    Baseline + Flipped & 300 & 90.33& 0.82 & 0.90\\
    \hline
    Baseline + Rotated & 400 & 95.25 & 0.91 &0.95\\
    \hline
  \end{tabular}
  }
\end{table}

From this it can be seen that each of the augmentation techniques have yielded significant accuracy improvements over just the baseline dataset.  The bigger the augmented dataset the greater the improvement, which is broadly in line with expectations.  The flipping and rotating augmentation techniques in particular gave results suggestive of very good performance, with accuracy levels 90 percent and above, with similarly good MCC and F-score metrics. 

\subsection{Pipeline Augmented Datasets}

Table \ref{tab:mbs32_all} summarises the results of the various pipelined experiments, as detailed in section \ref{meth:pipelined}.  For each dataset the model was evaluated using each of the three validation methods: 5 times cross-fold, 70\% split, and using the independent test dataset.  A mini-batch size was chosen of 32, which is fairly standard for datasets of this size and represents a good compromise between generalisation capabilities and computational efficiency.

The table shows the accuracy and MCC metrics for each permutation \footnote{The F-score metric was dropped for reasons of clarity in interpreting the table, and because it takes into account less information than the MCC statistic}. An average was taken for each measure against the validation methods as a way of combining the results into a single value. 

\begin{table}[htbp]
\caption{mini-batch size = 32}
\label{tab:mbs32_all}
\resizebox{\columnwidth}{!}{
\begin{tabular}{|*{10}{c|}}
\hline
\multicolumn{1}{|c}{\textbf{Dataset}} & \multicolumn{1}{|c}{\textbf{\#samples}} & \multicolumn{2}{|c}{\textbf{5x cross-fold}} & \multicolumn{2}{|c}{\textbf{70\% split}} & \multicolumn{2}{|c|}{\textbf{test dataset}} & \multicolumn{2}{|c|}{\textbf{Averaged}} \\
\cline{3-10}
& & \textbf{Acc} & \textbf{MCC} & \textbf{Acc} & \textbf{MCC}  & \textbf{Acc} & \textbf{MCC}& \textbf{Acc} & \textbf{MCC} \\
\hline
pipeline 1 & 800 & 98.00& 0.96&96.70 & 0.93&79.00& 0.60&90.23 & 0.83\\
\hline
pipeline 2 & 1200 & 99.50& 0.99& 98.60& 0.97& 75.00& 0.49& 91.03&0.82 \\
\hline
pipeline 3 & 2400 & 99.30& 0.98& 99.30& 0.99& 76.0& 0.54&91.53&0.84 \\
\hline
\end{tabular}
}
\end{table}

This table illustrates that in all cases that the trained model performs significantly better than the singly augmented datasets in their ability to correctly classify an image as containing/not containing garbage.  They do so with greater than 98\% accuracy when validated against the training data, and greater than 75\% against the independent test data.  This resulted in an overall average accuracy of around 90\% and MCC of greater than 0.82.  This represents strong performance in terms of classification accuracy and balanced predictions.

\section{Conclusion} \label{conc}

This study sought to investigate two issues:

\begin{itemize}
    \item
What are the most effective machine learning algorithms and techniques for developing a model that can accurately identify and classify small scale garbage dumps in the rural areas of Cyprus from satellite imagery?
\item 
To what extent can data augmentation techniques improve a machine learning
model for such small scale satellite image classification?
\end{itemize}

To answer these questions a baseline dataset of 100 training satellite image patches was collected using Google Earth \cite{google_corp_google_nodate}, which matched verified garbage dump locations.  For each garbage image a non garbage image from similar neighbouring terrain was collected. These images were correspondingly labelled as either ``garbage" or ``not garbage".  Several popular convolutional neural network implementations were trained and evaluated on this baseline set, with the ResNet-50 model showing the most promise.

The baseline dataset was then augmented by applying various combinations of sharpening, rotating, flipping, and cropping.  This resulted in ``pipelines" of 800, 1200 and 2400 sized image sets.  The ResNet-50 model, pretrained on the ImageNet dataset\cite{krizhevsky_imagenet_2017}, was further trained on each pipelined collection of images and the performance of each model evaluated with respect to the others.  

Various validation techniques were used to evaluate the accuracy and generalisation capabilities of the models, including cross-fold validation, employing a holdout dataset, and testing against a separate independently collected set of images.

The experiments showed that ResNet-50 provides an effective machine learning model in the detection of garbage dumps.  With just the small baseset of 100 images, a model could be trained that correctly predicts around 70\% of novel images.  When the size of this baseline sets was expanded using data augmentation then the predictive capabilities of the model increased dramatically, correctly classifying images in more than 90\% of cases. 

\subsection{Future Work}

Work is continuing in the area by:
\begin{itemize}
    \item Increasing the quantity and quality of the baseline dataset.  In time this will be made publicly available.
    \item Investigating automatic methods of obtaining appropriate satellite imagery.
    \item Developing a system that will allow the techniques explored to map the entire rural environment of Cyprus to develop a comprehensive ``garbage map".
\end{itemize}

\printbibliography
\end{document}